\documentclass[%
 reprint,
superscriptaddress,
 amsmath,amssymb,
 aps,
]{revtex4-2}
\usepackage{graphicx}% Include figure files
\usepackage{dcolumn}% Align table columns on decimal point
\usepackage{bm}% bold math
\usepackage[normalem]{ulem}
\usepackage[pdftex,dvipsnames,usenames]{xcolor}
\usepackage{color,soul}
\usepackage{bm}
\usepackage[dvipsnames]{xcolor}
\usepackage{esvect}
\usepackage{multirow}
\usepackage[normalem]{ulem}
\usepackage{nameref}
\usepackage{mathrsfs,soul}

\newcommand{\vis}[1]{\ensuremath{\mathcal{V}_{#1}}}
\newcommand{\vish}{\ensuremath{\mathcal{V}_H}}
\newcommand{\visv}{\ensuremath{\mathcal{V}_V}}
\newcommand{\h}[1]{\hat{#1}}
\newcommand{\ket}[1]{| #1 \rangle}
\newcommand{\bra}[1]{\langle #1 |}

\newcommand{\dm}{\hat{\rho}}
\newcommand{\I}{\mathscr{I}}

\newcommand{\avgpn}[1]{\langle #1 \rangle}
\renewcommand{\hat}{\widehat}

\begin{document}

\title{Quantum state tomography of undetected photons}

\author{Jorge Fuenzalida}
\email{jorge.fuenzalida@tu-darmstadt.de}
\affiliation{Institute for Quantum Optics and Quantum Information, Austrian Academy of Sciences, Boltzmanngasse 3, Vienna A-1090, Austria}
\affiliation{Vienna Center for Quantum Science and Technology (VCQ), Faculty of Physics, Boltzmanngasse 5, University of Vienna, Vienna A-1090, Austria}
\affiliation{Current address: Institute of Applied Physics, Technical University of Darmstadt, Schlo{\ss}gartenstraße 7, 64289 Darmstadt, Germany}

\author{Jaroslav Kysela}
\email{jaroslav.kysela@univie.ac.at}
\affiliation{Institute for Quantum Optics and Quantum Information, Austrian Academy of Sciences, Boltzmanngasse 3, Vienna A-1090, Austria}
\affiliation{Vienna Center for Quantum Science and Technology (VCQ), Faculty of Physics, Boltzmanngasse 5, University of Vienna, Vienna A-1090, Austria}

\author{Krishna Dovzhik}
\affiliation{Institute for Quantum Optics and Quantum Information, Austrian Academy of Sciences, Boltzmanngasse 3, Vienna A-1090, Austria}
\affiliation{Vienna Center for Quantum Science and Technology (VCQ), Faculty of Physics, Boltzmanngasse 5, University of Vienna, Vienna A-1090, Austria}

\author{Gabriela Barreto Lemos}
\affiliation{Instituto de F\'isica, Universidade Federal do Rio de
Janeiro, Av. Athos da Silveira Ramos 149,
Rio de Janeiro, CP: 68528, Brazil}

\author{Armin Hochrainer}
\affiliation{Institute for Quantum Optics and Quantum Information, Austrian Academy of Sciences, Boltzmanngasse 3, Vienna A-1090, Austria}
\affiliation{Vienna Center for Quantum Science and Technology (VCQ), Faculty of Physics, Boltzmanngasse 5, University of Vienna, Vienna A-1090, Austria}

\author{Mayukh Lahiri}
\email{mlahiri@okstate.edu}
\affiliation{Department of Physics, Oklahoma State University, Stillwater 74078, Oklahoma, USA}

\author{Anton Zeilinger}
\email{anton.zeilinger@univie.ac.at}
\affiliation{Institute for Quantum Optics and Quantum Information, Austrian Academy of Sciences, Boltzmanngasse 3, Vienna A-1090, Austria}
\affiliation{Vienna Center for Quantum Science and Technology (VCQ), Faculty of Physics, Boltzmanngasse 5, University of Vienna, Vienna A-1090, Austria}

\date{\today}

\begin{abstract}
The measurement  of quantum states is one of the most important problems in quantum mechanics. We introduce a quantum state tomography technique in which the state of a qubit is reconstructed, while the qubit remains undetected. 
The key ingredients are: (i) employing an additional qubit, (ii) aligning the undetected qubit with a known reference state by using path identity, and (iii) measuring the additional qubit to reconstruct the undetected qubit state. We theoretically establish and experimentally demonstrate the method with photonic polarization states. The principle underlying our method could also be applied to quantum entities other than photons.
\end{abstract}

\maketitle

\textit{Introduction.}---%
The process and interpretation of quantum measurements have been vastly studied since the early days of quantum mechanics \cite{neumann2013mathematische}. The results of measurements made on a given quantum system can be succinctly summarized in the form of a quantum state, which represents the complete description of the quantum system. The quantum state estimation spans over multiple techniques, such as projective measurements \cite{wheeler2014quantum,paris2004quantum}, weak measurements \cite{busch1996insolubility, brun2002simple,bergou2013extracting,solis2016experimental}, and device-independent measurements \cite{mayers1998quantum,colbeck2011private, agresti2019experimental}. The quantum state tomography consists in the identification of a quantum state by measurements on many identical copies of a quantum system \cite{nielsen2002quantum}. In the case of photons, the measurement is destructive via direct measurements~\cite{james2001measurement,altepeter2005photonic} or interferometric measurement \cite{sahoo2020quantum}. Furthermore, technical limitations might hinder an efficient detection process. For example, single-photon detectors are not equally efficient over the whole spectrum. Therefore, standard quantum tomography approaches may be challenging or even impossible in some cases.

Here, we introduce and implement a quantum tomography method in which the state of a single qubit is reconstructed without the qubit being detected or any direct measurement being performed on it~\footnote{We note here that for some quantum systems such as trapped ions, the quantum state can be determined by detecting fluorescence light, i.e., without  performing direct measurement on the ions~\cite{wineland1993ions,blatt2004bell}. However, such approaches are not applicable to photons and many other quantum entities like neutrons, electrons, etc.}. Instead, we introduce an additional particle (photon in our case). We then apply the concept of \emph{path identity} \cite{RevModPhys.94.025007} to build an interferometer, in which only the additional photon is detected via intensity measurement. We reconstruct the quantum state of the qubit from the interferometric data. Our method works for both pure and mixed states, and we experimentally demonstrate it for pure single-photon polarization states.

In contrast with existing tomographic methods, our method is applicable to cases in which the qubit cannot be detected for any technical or fundamental reason. This is because we do not detect the qubit and the additional photon can be chosen at a frequency for which adequate detectors are available.

\begin{figure}[ht]
%\centering
\includegraphics[width=1\linewidth]{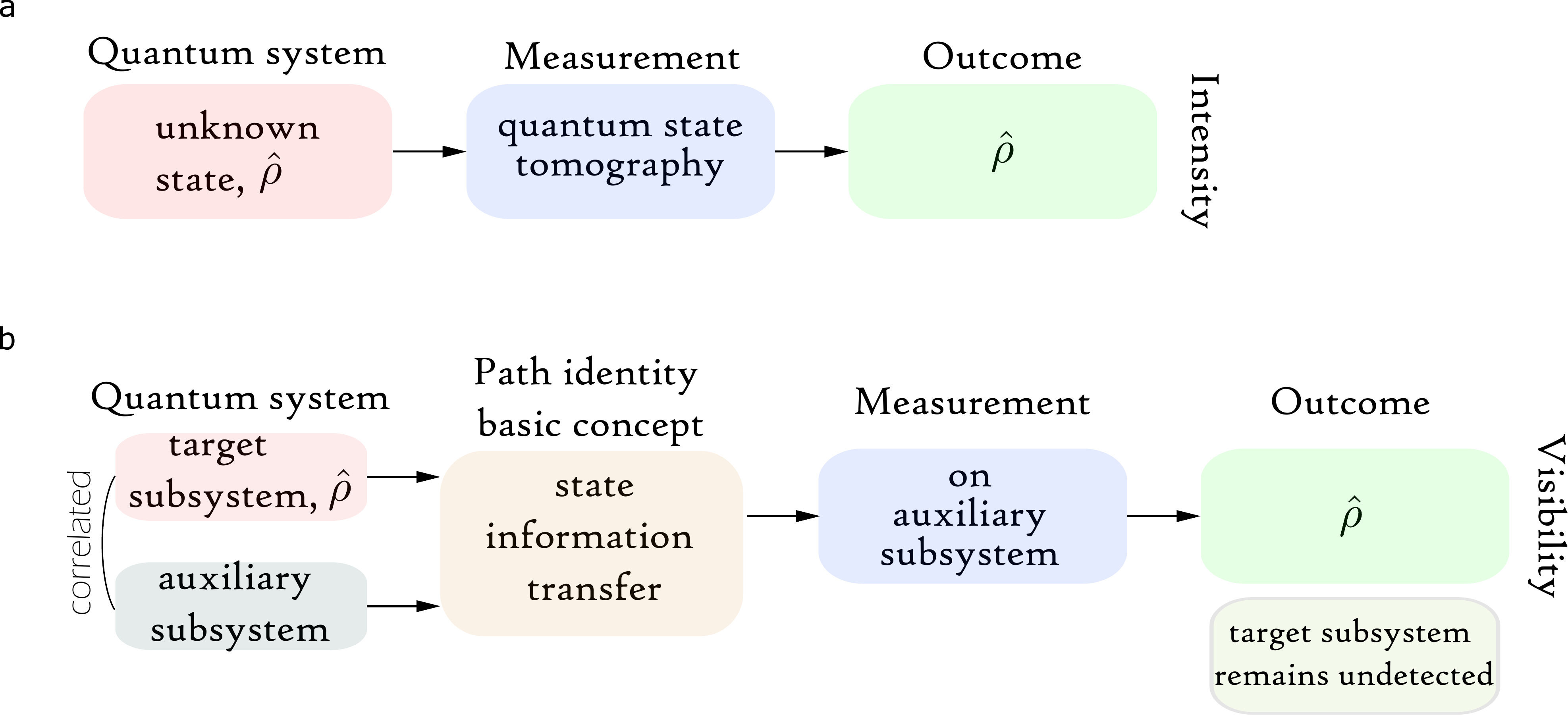}
\caption{ 
Quantum state tomography with detected and undetected particles. (a) Quantum state tomography retrieves the state of a quantum system by measuring an ensemble of identical particles. (b) In our scheme, the target subsystem for which we aim to obtain $\h{\rho}$ remains undetected. What is measured instead is an auxiliary subsystem that is quantum correlated to the target one.}
\label{fig:concept}
\end{figure}

Our method of quantum state tomography is based on the quantum phenomenon of induced coherence without induced emission \cite{zou1991induced,wang1991induced}, which can be viewed as interference by path identity \cite{RevModPhys.94.025007}. Mandel and collaborators first demonstrated this phenomenon using an interferometric arrangement consisting of two photon-pair sources. They induced coherence between two modes of one photon by making paths of its partner photon \emph{identical}. They obtained a single-photon interference pattern detecting the former photon only. The detection of the partner photon was not required. We henceforth refer to this phenomenon as ``induced coherence'' for simplicity. The potential of induced coherence has already been recognised as it has found use in quantum imaging~\cite{lemos_quantum_2014}, spectroscopy~\cite{kalashnikov2016infrared}, holography~\cite{topfer2022quantum}, sensing~\cite{kutas2020terahertz}, optical coherence tomography~\cite{valles2018optical,paterova2018tunable}, measurement of two-photon correlations and entanglement~\cite{lahiri2017twin,hochrainer2017quantifying,lahiri2021characterizing,lemos2023one}. This phenomenon has also been exploited to investigate the interplay between the polarization and coherence \cite{lahiri2017partial,paterova2019polarization}. Nevertheless, to date no technique based on induced coherence has been developed to fully reconstruct the polarization state of single photons.

\begin{figure}[ht]
%\centering
\includegraphics[width=1\linewidth]{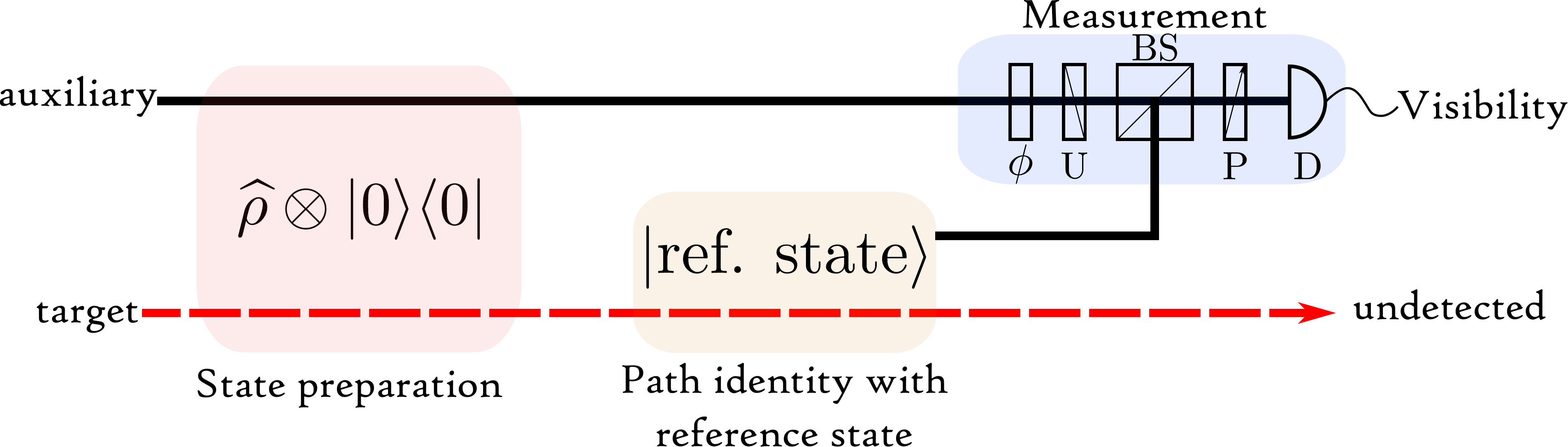}
\caption{Path identity concept of implementation. The crucial point in our scheme is transferring the state information from the target to the auxiliary particle which is done by a path identity approach. Measurements on the auxiliary particle allow us to fully retrieve $\h{\rho}$ of the target particle. $\phi$: phase shifter, U: unitary operation, P: projector, BS: beam splitter, D: detector.
}
\label{fig:implem.}
\end{figure}

\textit{Principle.}---%
\label{sec:principle}% %
In the standard polarization quantum state tomography [Fig.~\ref{fig:concept}(a)] a photon is prepared in a general state $\h{\rho}$ and then subjected to intensity measurements performed in various settings given by the Pauli operators. The recorded results are then used to obtain the Stokes parameters and the associated density matrix $\h{\rho}$ \cite{james2001measurement}. In our technique [Fig.~\ref{fig:concept}(b)], the target photon whose state $\h{\rho}$ we reconstruct remains undetected. What is measured instead is an auxiliary photon with which the target photon interacts. Information about the state $\h{\rho}$ is transferred to the auxiliary photon by employing a path identity approach [Fig.~\ref{fig:implem.}]. Through this step, the state of the target photon changes, and its representation can be found in the Supplemental Material. To obtain $\h{\rho}$, the auxiliary photon is prepared, projected, and detected in different configurations. The state $\h{\rho}$ is retrieved from the interference patterns obtained by measuring the auxiliary photon.

\textit{Method.}---%
\label{sec:2-mode-2}%
\begin{figure}[ht]
%\centering
\includegraphics[width=1\linewidth]{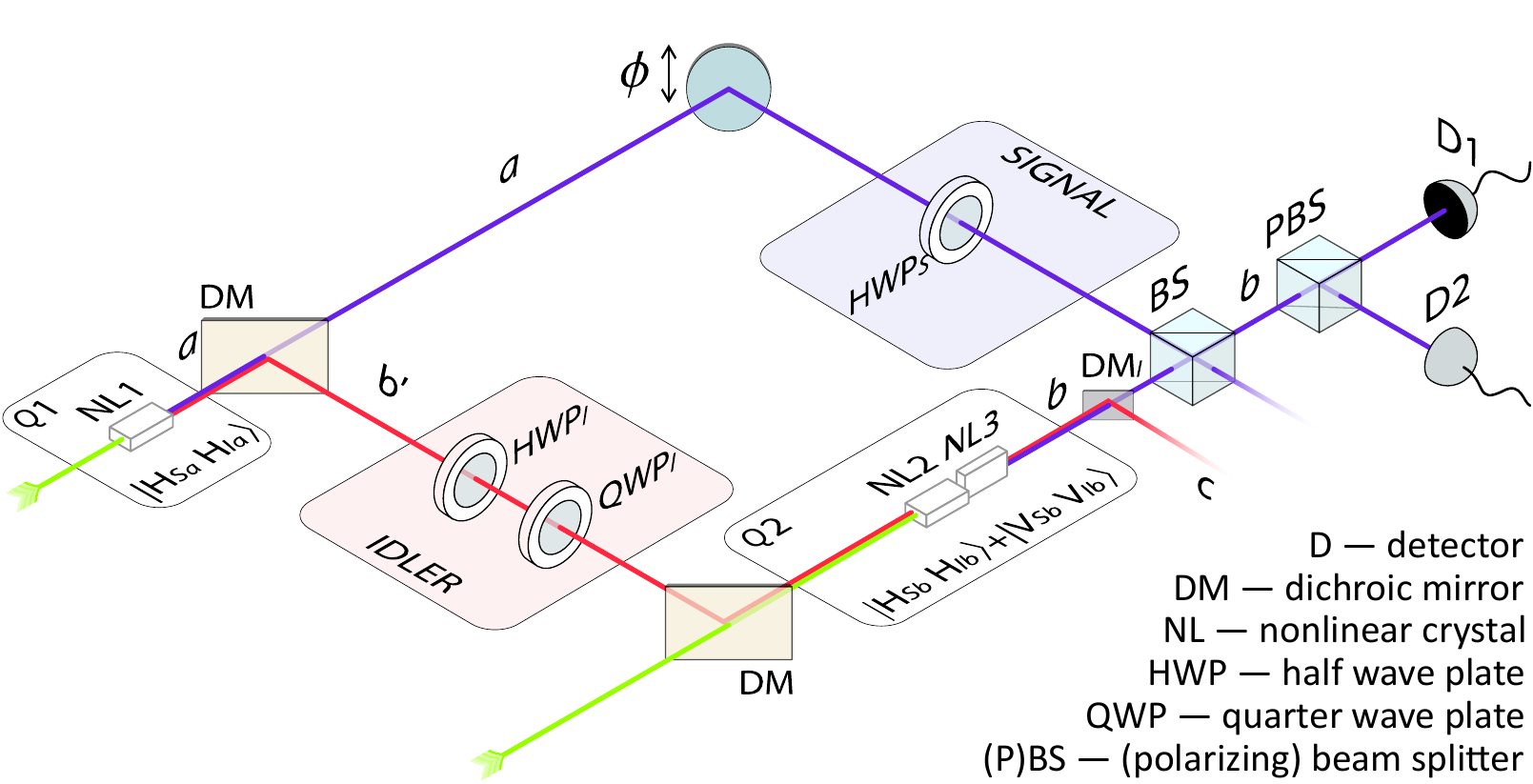}
\caption{The experimental setup for the proposed quantum tomography method of polarization states. The pump, the idler and the signal photons are represented by the green, red and blue lines, respectively. One photon pair is created in a coherent superposition of three nonlinear crystals (NL1--NL3). Idler beams are overlapped in order to erase the which-source information, the lack of which induces coherence of the signal photon. Signal photons emitted in each crystal are superposed on a beam splitter (BS), split into $H$ and $V$ polarization by a polarizing beam splitter (PBS) and detected by detectors D1 and D2. The recorded intensities of the signal photon exhibit interference as the independent phase $\phi$ is varied. Our method allows one to identify the polarization state of the idler photon emitted in crystal NL1 (source Q1) even though it is never detected. $a$, $b$, $b'$, and $c$ stand for optical paths.}
\label{fig:main-setup}
\end{figure}
Our aim is to reconstruct the quantum state of a photon, henceforth referred to as the idler photon, whose density matrix has the general form
\begin{equation}
    \dm_I = \begin{pmatrix}
    P_H & \I \sqrt{P_H P_V} e^{-i\xi} \\
    \I \sqrt{P_H P_V} e^{i \xi} & P_V
    \end{pmatrix}.
    \label{eq:rhoI}
\end{equation}
In the formula above, $P_H$ ($P_V$) is the probability of detecting the horizontal $H$ (vertical $V$) polarization of the idler photon, $P_H + P_V = 1$; $0 \leq \xi < 2 \pi$ is a relative phase between $H$ and $V$; and $0 \leq \I \leq 1$ quantifies the purity of the state. It can be readily checked that $\I=0$ and $\I=1$ correspond to fully mixed and pure states, respectively. 
\par
Our reconstruction technique employs the experimental setup depicted in Fig.~\ref{fig:main-setup}. We employ an additional photon, referred to as the signal photon, such that the two photons are in the state
\begin{align}
\dm_{1}=\ket{H_{Sa}}\bra{H_{Sa}} \otimes \dm_I,
\label{eq:rho1}
\end{align}
where $\ket{H_{S_a}}$ represents a horizontally polarized signal photon propagating along path $a$, and the idler photon $I$ propagates along path $b'$ in the state $\dm_I$. Our technique uses two sources of photon pairs, denoted by Q1 and Q2. The first source comprises a single nonlinear crystal and produces photons in a separable state. After the state preparation of the idler photon, this state is given by Eq.~\eqref{eq:rho1}. The second source comprises two nonlinear crystals in a cross-crystal configuration and produces photon pairs in a maximally entangled state $\dm_2 = \ket{\psi_2} \bra{\psi_2}$ in path $b$, where
\begin{align}
\ket{\psi_2} = \frac{1}{\sqrt{2}}(\ket{H_{Sb}}\ket{H_{Ib}}+\ket{V_{Sb}}\ket{V_{Ib}}).
\end{align}
This state plays the role of a reference state of polarization, with respect to which state $\dm_I$ is probed, as explained below.

$Q_1$ and $Q_2$ are pumped coherently. The pump power for both sources is low, such that the two sources do not emit photons simultaneously and the effect of stimulated emission can be neglected. In this case, one photon pair is in a coherent superposition of being emitted by the sources Q1 and Q2. The density matrix representing a photon pair in such a scenario is given by Eq.~(4) of the Supplemental material.

The alignment of photons' paths plays a crucial role in our technique. We superimpose the two signal-photon paths $a$ and $b$ at a beam splitter and later the two polarization modes in path $b$ are separated by a polarizing beam splitter. The final $H$ ($V$) component is collected by detector D1 (D2); see Fig.~\ref{fig:main-setup}. When the idler beam emitted from the source Q1 is aligned precisely such that it overlaps with the idler beam emitted from the source Q2, the source information is erased, and an interference pattern arises in intensities detected by D1 and D2. Therefore, we have an effective two-fold induced coherence in two orthogonal polarization modes $H$ and $V$, where the level of indistinguishability depends on $\dm_I$.

When the pump powers are adjusted appropriately (see Supplemental material), the intensity (photon counting rate) for $H$ polarization at detector D1 is given by
\begin{equation}
    \avgpn{\hat{R}_{H}} = \frac{1}{4} \left( 1+|T_H|\sqrt{P_H}\cos(\phi) \right),
    \label{eq:RH}
\end{equation}
provided that the half-wave plate HWP$_S$ is set to $0^\circ$, where $T_H$ is the total amplitude transmission coefficient of the $H$ polarization, which takes into account all optical components placed on path $b'$, and $\phi \in \mathbb{R}$ is the relative phase between the two sources. The rate $\avgpn{\hat{R}_{V}}$ for vertical polarization stays constant for such a setting. To record the interference pattern for $V$ polarization, we set the HWP$_S$ to $45^\circ$, such that the signal photon's state in path $a$ turns to $\ket{V_{Sa}}$. In this case, rate $\avgpn{\hat{R}_{H}}$ monitored by D1 stays constant, while detector D2 records photon counts whose rate varies according to
\begin{equation}
    \avgpn{\hat{R}_{V}} = \frac{1}{4} \left( 1+\I|T_V|\sqrt{P_V}\cos(\phi+\xi) \right),
    \label{eq:RV}
\end{equation}
where $T_V$ is the transmission coefficient for the $V$ polarization, defined analogously to $T_H$, and $\xi$ was introduced in Eq.~\eqref{eq:rhoI}. Upon varying $\phi$ the two expressions above display interference with visibilities 
\footnote{The visibility is $\mathcal{V}=(\avgpn{R_{\mathrm{max}}}-\avgpn{R_{\mathrm{min}}})/(\avgpn{R_{\mathrm{max}}}+\avgpn{R_{\mathrm{min}}})$, where $\avgpn{R_{\mathrm{max}}}$ and $\avgpn{R_{\mathrm{min}}}$ are the maximum and minimum intensities, respectively.}
\begin{eqnarray}
    \vis{H} & = & |T_H|\sqrt{P_H}, \label{eq:visH} \\
    \vis{V} & = & \I |T_V|\sqrt{P_V}. \label{eq:visV} 
\end{eqnarray}
These formulas quantify the level of indistinguishability in the two orthogonal polarization modes. We can perform a separate measurement of the amplitude transmission coefficients $|T_H|$ and $|T_V|$ (i.e., losses and alignment). These coefficients determine the maximum achievable visibilities. The visibility $\vis{H}$ allows us to infer the value of $P_H$ and $P_V = 1 - P_H$, the visibility $\vis{V}$ contains information about the degree of coherence $\I$ and the local phase $\xi$ is equal to the relative phase shift between interference fringes for $\avgpn{\hat{R}_{H}}$ \eqref{eq:RH} and $\avgpn{\hat{R}_{V}}$ \eqref{eq:RV}. This way we can reconstruct state $\dm_I$ in Eq.~\eqref{eq:rhoI}.

\begin{figure*}[htbp]
\centering
\includegraphics[width=.95\linewidth]{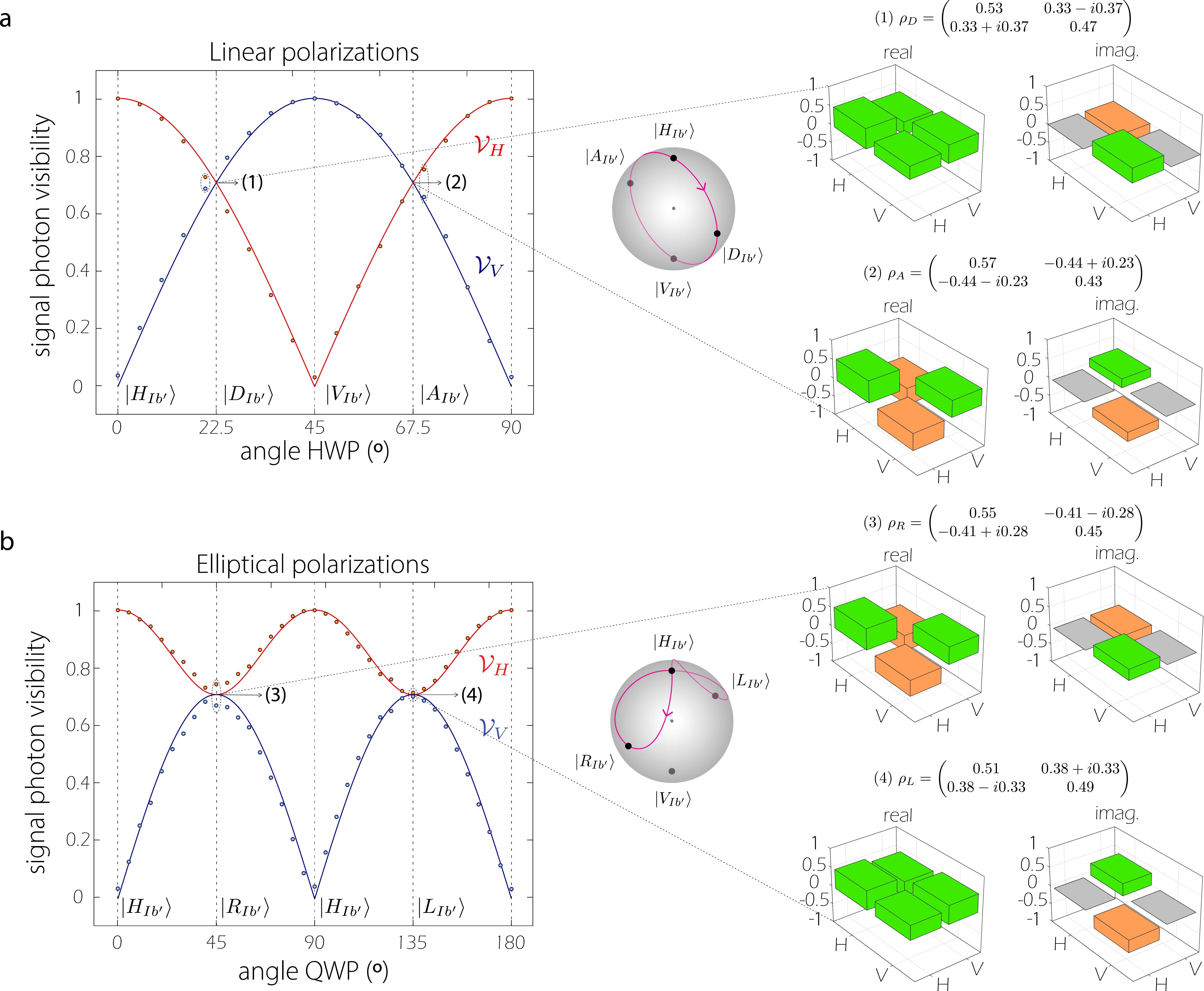}
\caption{Experimental results for pure state tomography of photons that are never detected. Main results of visibility measurements are shown in the left panels. The red and blue colors represent the horizontal and vertical polarization, respectively. (a) The idler photon is prepared in different linear polarization states by HWP$_I$, see the Bloch sphere in the inset. The signal photon visibility for each linear polarization is plotted against the angle of the plate HWP$_I$. (b) The idler photon is prepared in different elliptical polarization states by QWP$_I$, starting in $\ket{H_{Ib'}}$. The signal photon visibility is plotted against the angle of the plate QWP$_I$. The calibrated experimental data points in (a) and (b) are in agreement with the theoretical predictions, which are given by the solid lines. The idler states $\ket{D_{Ib'}}$ and $\ket{A_{Ib'}}$, designated by (1) and (2) in (a), and $\ket{R_{Ib'}}$ and $\ket{L_{Ib'}}$, designated by (3) and (4) in (b), are all represented by the same visibilities of the signal photon. Nevertheless, these four states have different relatives phases. We have included the experimental density matrices obtained for these four states in the insets (1)--(4).}
\label{Fig:main-results}
\end{figure*}

\textit{Implementation.}---%
\label{sec:experiment}%
We demonstrated the described technique of reconstruction of the idler state experimentally. We used ppKTP crystals emitting photon pairs by spontaneous parametric down-conversion in a non-degenerate, type-0 process. Signal (idler) photons had a central wavelength of 842 (780)~nm. In the detection part, we used a 800~nm long-pass filter and a $842 \pm 1$~nm interference filter to select only signal photons.

The tomography procedure needs to be calibrated to compensate for the misalignment in the experimental implementation. To that end, we measured the maximum visibilities $T_H$ and $T_V$ that can be achieved experimentally for $H$ and $V$ polarization. We experimentally obtained $T_H$ = 0.85 $\pm 0.03$ and $T_V$ = 0.73 $\pm 0.02$ \footnote{The difference between these maximum visibilities is caused by the impossibility of imaging the crystal plane of the source Q1 simultaneously to the crystal planes of the two crystals comprising the source Q2. However, this difference could be fixed by building the source Q2 in the Sagnac geometry.}. The calibration itself consisted in the division of each experimentally obtained visibility $\vish$ ($\visv$) by $T_H$ ($T_V$).

A cascade of a half-wave plate (HWP$_I$) and a quarter-wave plate (QWP$_I$) is used to prepare the polarization state $\dm_I$ of the idler photon generated by the source Q1 in the path $b'$. The tomography was performed by taking measurements in two configurations of the signal photon, $\ket{H_{Sa}}$ and $\ket{V_{Sa}}$, which were prepared by the plate HWP$_S$. For the configuration $\ket{H_{Sa}}$ ($\ket{V_{Sa}}$) we recorded the intensity $\avgpn{\hat{R}_{H}}$ ($\avgpn{\hat{R}_{V}}$) at the detector D1 (D2) while the phase $\phi$ was varied. Fitting a sinusoidal function to the recorded intensity profiles we extracted the visibilities for the signal photon as well as the phase shift for each state. The values of $P_H$, $P_V$, and $\xi$ were then extracted using Eqs.~\eqref{eq:RH} and \eqref{eq:RV}. Additionally, we implemented a maximum likelihood estimation algorithm to reconstruct density matrices from our data~\footnote{Additional information about the explicit form and evolution of the photons' state, polarization measurement, calibration, state reconstruction, post-measurement state, and measured fidelities can be found in the supplemental material.}.

Experimental results are presented in Fig.~\ref{Fig:main-results}, where dots represent experimental data, whereas solid lines represent theoretical prediction. We performed two measurement scans. In the first scan the entire range of pure linearly polarized states of the idler photon was traversed by rotating HWP$_I$. Calibrated experimental results, which are presented in Fig.~\ref{Fig:main-results} (a), agree well \footnote{The standard deviation is obtained from sinusoidal function fitted to the experimental data.} with the theoretical predictions given by Eqs.~\eqref{eq:visH} and \eqref{eq:visV}. Indeed, initially $\vish$ is maximal and $\visv$ is minimal as the idler state is $H$-polarized. As HWP$_I$ is rotated, $\vish$ ($\visv$) gradually decreases (increases) until it reaches its minimum (maximum) when the idler photon is $V$-polarized. The mean fidelity of states reconstructed in this scan is found to be $\Bar{F}_{\mathrm{HWP}}=0.97$.

In the second scan, QWP$_I$ was rotated to generate a series of pure elliptically polarized states ranging from the right-circular polarization to the horizontal polarization and then further to the left-circular polarization. We started with $\ket{H_{Ib'}}$. As QWP$_I$ is rotated and the state of the idler photon acquires a complex $V$-polarized component, the visibility for $H$ decreases whereas that of $V$ increases until they both reach the extreme value of $1/\sqrt{2}$ when the state is right- or left-circular. This behavior is closely followed by the experimental data, which are presented in Fig.~\ref{Fig:main-results} (b). The states measured in this scan have a mean fidelity of $\Bar{F}_{\mathrm{QWP}}=0.92$.

As has been pointed out in the previous section, the visibilities $\vish$ \eqref{eq:visH} and $\visv$ \eqref{eq:visV} do not identify the idler photon's state unambiguously. Any two states $\dm_I$ and $\dm'_I$ for which $\vish = \vish'$ and $\visv = \visv'$ will be misidentified. For precise identification also the phase shift between interference patterns in Eqs.~\eqref{eq:RH} and \eqref{eq:RV} is needed. This fact is illustrated for states $\ket{D_{Ib'}}$ and $\ket{A_{Ib'}}$ (diagonal and anti-diagonal) in the insets of Fig.~\ref{Fig:main-results} (a), and states $\ket{R_{Ib'}}$ and $\ket{L_{Ib'}}$ (right- and left-circular) in the insets of Fig.~\ref{Fig:main-results} (b). Visibilities $\vish$ as well as $\visv$ are identical for all these states, but the relative phases differ.

The experimental results clearly demonstrate the feasibility of our reconstruction technique, as long as pure polarization states of photons are considered. We successfully characterized all the linearly polarized states and also a specific class of elliptically polarized states. Fidelities of all the states, shown in a plot in the Supplemental Material, lie above $F_{\mathrm{min}}=0.72$.

\textit{Conclusion.}---%
\label{sec:conclusion}
In this paper we introduced a novel tomography method that allows one to determine a polarization state of a photon that is never detected. This photon is a part of one photon pair that is emitted in a coherent superposition of two spatially separated sources. The first source emits idler photons in an unknown state that is to be reconstructed. The second source produces a reference state that contains two orthogonal polarization modes~\footnote{We note that the crystal NL3 can in principle be replaced by additional retarder plates placed in the idler path before NL2.}. As a result of the indistinguishability in the two polarization modes the resulting interference patterns contain all the information about the idler photon that allows us to fully reconstruct its state. We demonstrated experimentally this reconstruction for two classes of idler states---linearly polarized states and specifically elliptically polarized states. We also presented the theoretical treatment of the general case when the idler photon's polarization state is mixed. The fidelities of states can be increased by improving the phase stability in our experiment.

The main advantage over traditional techniques stems from the fact that in our approach only signal photons are detected and these can have a different wavelength from that of the idler photons. In certain frequency (wavelength) ranges, efficient detectors are not affordable or available. In our approach, the high flexibility of the SPDC process allows us to choose the favorable wavelength that can be efficiently detected \cite{kviatkovsky2020microscopy, paterova2020hyperspectral,kutas2020terahertz,paterova2020quantum}.

To conclude, this work provides a new tool for the precise identification of single-photon states with high potential in a wide range of applications. Our approach can be adapted to the tomography of high-dimensional degrees of freedom of single photons, such as the orbital angular momentum \cite{erhard2020advances}. Including spatial modes in our technique, i.e. position and momentum, one can implement polarization quantum imaging with undetected photons \cite{lemos_quantum_2014}. It can also be used in the characterization of birefringent materials, sensing, and polarization-sensitive optical coherence tomography~\cite{valles2018optical,paterova2018tunable}. Note that the post-measurement state of the idler photon still carries partial information about the state preparation. If the idler detection is not a problem, our scheme can be used in quantum communication protocols between two parties, where one party obtains information through interference patterns (signal photon), and the other party obtains probabilistic information via projective measurements (idler photon). It has also not escaped our notice that our method can be adapted to any two-level quantum system, such as atoms, ions, or molecules.

JF thanks G. Carvacho for helpful discussions. The authors thank R. Blach for technical support. This work was supported by the Austrian Academy of Sciences (ÖAW). JF also acknowledges ANID for the financial support (Becas de doctorado en el extranjero “Becas Chile”/ 2015 – No. 72160487). JK acknowledges the support of the University of Vienna via the project QUESS (Quantum Experiments on Space Scale). KD acknowledges the support of the Austrian Science Fund (FWF): F40 (SFB FoQuS).

\nocite{*}

\bibliography{main}

\end{document}

\title{Quantum state tomography of undetected photons: Supplemental material}

\author{Jorge Fuenzalida}
\email{jorge.fuenzalida@tu-darmstadt.de}
\affiliation{Institute for Quantum Optics and Quantum Information, Austrian Academy of Sciences, Boltzmanngasse 3, Vienna A-1090, Austria}
\affiliation{Vienna Center for Quantum Science and Technology (VCQ), Faculty of Physics, Boltzmanngasse 5, University of Vienna, Vienna A-1090, Austria}
\affiliation{Current address: Institute of Applied Physics, Technical University of Darmstadt, Schlo{\ss}gartenstraße 7, 64289 Darmstadt, Germany}

\author{Jaroslav Kysela}
\email{jaroslav.kysela@univie.ac.at}
\affiliation{Institute for Quantum Optics and Quantum Information, Austrian Academy of Sciences, Boltzmanngasse 3, Vienna A-1090, Austria}
\affiliation{Vienna Center for Quantum Science and Technology (VCQ), Faculty of Physics, Boltzmanngasse 5, University of Vienna, Vienna A-1090, Austria}

\author{Krishna Dovzhik}
\affiliation{Institute for Quantum Optics and Quantum Information, Austrian Academy of Sciences, Boltzmanngasse 3, Vienna A-1090, Austria}
\affiliation{Vienna Center for Quantum Science and Technology (VCQ), Faculty of Physics, Boltzmanngasse 5, University of Vienna, Vienna A-1090, Austria}

\author{Gabriela Barreto Lemos}
\affiliation{Instituto de F\'isica, Universidade Federal do Rio de
Janeiro, Av. Athos da Silveira Ramos 149,
Rio de Janeiro, CP: 68528, Brazil}

\author{Armin Hochrainer}
\affiliation{Institute for Quantum Optics and Quantum Information, Austrian Academy of Sciences, Boltzmanngasse 3, Vienna A-1090, Austria}
\affiliation{Vienna Center for Quantum Science and Technology (VCQ), Faculty of Physics, Boltzmanngasse 5, University of Vienna, Vienna A-1090, Austria}

\author{Mayukh Lahiri}
\email{mlahiri@okstate.edu}
\affiliation{Department of Physics, Oklahoma State University, Stillwater, Oklahoma, USA}

\author{Anton Zeilinger}
\email{anton.zeilinger@univie.ac.at}
\affiliation{Institute for Quantum Optics and Quantum Information, Austrian Academy of Sciences, Boltzmanngasse 3, Vienna A-1090, Austria}
\affiliation{Vienna Center for Quantum Science and Technology (VCQ), Faculty of Physics, Boltzmanngasse 5, University of Vienna, Vienna A-1090, Austria}

\maketitle

\section{The total state of a photon pair emitted from sources Q1 and Q2}

To study the state of photons produced by the setup we apply the analysis developed in Ref. \cite{lahiri2021characterizing}.
The most general state of two photons that can be generated by sources Q1 and Q2 (after state preparation) discussed in the main text is of the form 
\begin{multline}
\dmt = |b_1|^2 \ \ket{H_{Sa}} \bra{H_{Sa}} \otimes \dm_I + |b_2|^2 \dm_2 \\ + \left[b_1 b_2^* \text{(cross terms)} + \text{h.c.}\right],
\label{eq:rho_tot}
\end{multline}
where $\dm_I$ is the unknown state of the idler photon given by Eq.~(2) in the main text; $b_1$ and $b_2$ are weighting factors that represent the unbalanced pumping of the two sources ($|b_1|^2+|b_2|^2=1$); and $\dm_2 = \ket{\psi_2} \bra{\psi_2}$ is the state of the second source with
\begin{align}
\ket{\psi_2} = \sqrt{P_{H2}} \ket{H_{Sb}}\ket{H_{Ib}} + e^{i \theta} \sqrt{P_{V2}} \ket{V_{Sb}}\ket{V_{Ib}}.
\end{align}
To express the total state $\dmt$ in the explicit matrix form, let us choose the following basis
\begin{multline}
    \{ \ket{H_{Sa} H_{Ib'}}, \ket{H_{Sa} V_{Ib'}}, \ket{V_{Sa} H_{Ib'}}, \ket{V_{Sa} V_{Ib'}}, \\
    \ket{H_{Sb} H_{Ib}}, \ket{H_{Sb} V_{Ib}}, \ket{V_{Sb} H_{Ib}}, \ket{V_{Sb} V_{Ib}} \},
\end{multline}
where $a$, $b$, and $b'$ denotes paths in the setup as given in Fig.~3 in the main text.

In this basis, the total state reads
\begin{widetext}
\begin{equation}
\dmt = 
    \left(
\begin{array}{cccccccc}
 % 1st line
 \left| b_1 \right|^2 P_H & \left| b_1\right|^2 \I \sqrt{P_H P_V} \, e^{-i \xi} 
   & 0 & 0 & b_1 b_2^* \sqrt{P_H P_{H2}} & 0 & 0 & b_1 b_2^* \sqrt{P_H P_{V2}} \, e^{-i \theta} \\
 % 2nd line
 \left| b_1\right|^2 \I \sqrt{P_H P_V} \, e^{i \xi} & \left| b_1 \right|^2 P_V & 0 & 0 & b_1 b_2^* \J \sqrt{P_V P_{H2}} \, e^{i \xi} & 0 & 0 & b_1 b_2^* \J' \sqrt{P_V P_{V2}} \, e^{-i (\theta-\xi)} \\
 % 3rd line
 0 & 0 & 0 & 0 & 0 & 0 & 0 & 0 \\
 % 4th line
 0 & 0 & 0 & 0 & 0 & 0 & 0 & 0 \\
 % 5th line
 b_1^* b_2 \sqrt{P_H P_{H2}} & b_1^* b_2 \J \sqrt{P_V P_{H2}} \, e^{-i \xi} & 0 & 0 & \left| b_2 \right|^2
   P_{H2} & 0 & 0 & \left| b_2 \right|^2 \sqrt{P_{H2} P_{V2}} \, e^{-i \theta} \\
 % 6th line
 0 & 0 & 0 & 0 & 0 & 0 & 0 & 0 \\
 % 7th line
 0 & 0 & 0 & 0 & 0 & 0 & 0 & 0 \\
 % 8th line
 b_1^* b_2 \sqrt{P_H P_{V2}} \, e^{i \theta} & b_1^* b_2 \J' \sqrt{P_V P_{V2}} \, e^{i (\theta - \xi)} & 0 & 0 & \left|
   b_2 \right|^2 \sqrt{P_{H2} P_{V2}} \, e^{i \theta} & 0 & 0 & \left| b_2 \right|^2 P_{V2}  \\
\end{array}
\right).
\label{eq:rho_matrix}
\end{equation}
\end{widetext}
Parameters $\J$ and $\J'$ quantify the coherence between the $V$ polarization of the first source and the $H$ and $V$ polarization of the second source, respectively. In order for matrix $\dmt$ in Eq.~\eqref{eq:rho_matrix} to be a valid density operator, it must have a unit trace and be positive semidefinite. The former condition is easy to check, for the latter condition one can use the generalization of Sylvester's criterion for positive semidefinite matrices~%[J. Guid. Control. Dyn. \textbf{9}, 121 (1986)].  
\cite{sylvCrit}. 
The requirement of positive semidefiniteness implies that $0 \leq \I = \J = \J' \leq 1$.

\section{Evolution of the total state}

We model the alignment process as an effective beam splitter that deflects with probability $|R_H|^2$ ($|R_V|^2$) the $H$-polarized ($V$-polarized) idler beam propagating along path $b'$ from source Q1 into auxiliary path $w$. With the complementary probability $|T_H|^2$ ($|T_V|^2$) the effective beam splitter directs the idler beam into path $b$. That is,
\begin{eqnarray}
    \ket{H_{Ib'}} & \to & R_H \ket{H_{Iw}} + T_H \ket{H_{Ib}}, \\
    \ket{V_{Ib'}} & \to & R_V \ket{V_{Iw}} + T_V \ket{V_{Ib}},
\end{eqnarray}
where $|R_H|^2 + |T_H|^2 = 1$ and $|R_V|^2 + |T_V|^2 = 1$. These transformation rules are applied to the total state $\dmt$ in Eq.~\eqref{eq:rho_tot}.

After the alignment of the idler beams we only manipulate the signal photon in the rest of the setup. For reasons that are explained later on, we use two settings of the signal photon's polarization. One, when its polarization in path $a$ is prepared in state $\ket{H_{Sa}}$, see Eq.~\eqref{eq:rho_tot}, and one, when it is rotated with a half-wave plate into the vertical polarization $\ket{V_{Sa}}$. When the first setting is considered, the reduced density operator of the signal photon in basis $\{ \ket{H_{Sa}}, \ket{V_{Sa}}, \ket{H_{Sb}}, \ket{V_{Sb}} \}$ is given by
\begin{equation}
\dm_S = 
    \left(
\begin{array}{cccccccc}
 % 1st line
 \left| b_1 \right|^2 & 0 & \rho_{12} & \rho_{14} \\
 % 2nd line
 0 & 0 & 0 & 0 \\
 % 3rd line
 \rho_{12}^* & 0 & \left| b_2 \right|^2 P_{H2} & 0 \\
 % 4th line
 \rho_{14}^* & 0 & 0 & \left| b_2 \right|^2 P_{V2}
\end{array}
\right),
\label{eq:rho_S}
\end{equation}
where the off-diagonal terms read $\rho_{12} = T_H b_1 b_2^* \sqrt{P_H P_{H2}}$ and $\rho_{14} = T_V b_1 b_2^* \I \, \sqrt{P_V P_{V2}} \, e^{i(\xi-\theta)}$. In the second case, the reduces density operator takes the form
\begin{equation}
\dm'_S = 
    \left(
\begin{array}{cccccccc}
 % 1st line
 0 & 0 & 0 & 0 \\
 % 2nd line
 0 & \left| b_1 \right|^2 & \rho_{12} & \rho_{14} \\
 % 3rd line
 0 & \rho_{12}^* & \left| b_2 \right|^2 P_{H2} & 0 \\
 % 4th line
 0 & \rho_{14}^* & 0 & \left| b_2 \right|^2 P_{V2}
\end{array}
\right).
\label{eq:rho_Sp}
\end{equation}
After adjusting the polarization of the signal photon in path $a$, the signal beams from source Q1 and Q2 are superimposed at a beam splitter. The action of the beam splitter can be represented by formulas
\begin{equation}
    \dm_S \to \mathrm{BS} \cdot \dm_S \cdot \mathrm{BS}^\dagger, \quad \dm'_S \to \mathrm{BS} \cdot \dm'_S \cdot \mathrm{BS}^\dagger,
\end{equation}
where BS is a matrix of the following form 
\begin{equation}
\mathrm{BS} = 
\frac{1}{\sqrt{2}} \, \begin{pmatrix}
1 & 1 \\ 1 & -1
\end{pmatrix} \, \otimes \,
\begin{pmatrix}
1 & 0 \\ 0 & 1
\end{pmatrix}.
\end{equation}

\section{Polarization measurement}

In the end, we perform standard polarization measurement in the $H/V$ basis of signal photons leaving the beam splitter along path $b$. The resulting photon count rates in the $H$ component for state $\dm_S$ in Eq.~\eqref{eq:rho_S} are given by
\begin{equation}
    \avgpn{\hat{R}_{H}} = \bra{H_{Sb}} \, \mathrm{BS} \cdot \dm_S \cdot \mathrm{BS}^\dagger \, \ket{H_{Sb}},
\end{equation}
and analogously for the $V$ component and state $\dm'_S$. In the experiment, we use two measurement settings. For the first setting, when the signal photon's polarization in path $a$ is equal to $\ket{H_{Sa}}$, we obtain the following photon count rates. For the $H$ component we get
\begin{multline}
     \avgpn{\hat{R}_{H}} = \frac{1}{2} \Big( \left| b_1 \right|^2+\left| b_2 \right|^2 P_{H2} \\ + 2 \left| b_1 \right| \left| b_2 \right| \left| T_H \right|  \sqrt{P_H P_{H2}} \cos (\phi) \Big),
     \label{eq:RHs}
\end{multline}
where $b_1 \in \mathbb{R}$ and $b_2 = |b_2|e^{i \phi}$. The visibility of this expression, when $\phi$ is varied, equals
\begin{equation}
    \vis{H} = \frac{2 \left| b_1 \right| \left| b_2 \right| \left| T_H \right|  \sqrt{P_H P_{H2}}}{\left| b_1 \right|^2+\left| b_2 \right|^2 P_{H2}}.
\end{equation}
At the same time, the $V$ component stays constant and equals $\avgpn{\hat{R}_{V}} = |b_2|^2 P_{V2}/2$. For the second setting we rotate the polarization of signal photons in path $a$, such that their state is $\ket{V_{Sa}}$. The count rate for $H$ component now stays at a constant value of $\avgpn{\hat{R}_{H}} = |b_2|^2 P_{H2}/2$. The $V$ component, on the contrary, varies like 
\begin{multline}
    \avgpn{\hat{R}_{V}} = \frac{1}{2} \Big(\left| b_1 \right|^2+\left| b_2 \right|^2 P_{V2} \\ + 2 \, \I \, \left| b_1 \right| \left| b_2 \right| \left| T_V \right| \sqrt{P_V P_{V2}} \cos (\phi + \theta - \xi) \Big).
    \label{eq:RVs}
\end{multline}
The visibility of this expression is equal to
\begin{equation}
    \vis{V} = \frac{2 \, \I \, \left| b_1 \right| \left| b_2 \right| \left| T_V \right| \sqrt{P_V P_{V2}}}{\left| b_1 \right|^2+\left| b_2 \right|^2 P_{V2}}.
\end{equation}

Our experimental arrangement corresponds to conditions $b_2 = \sqrt{2} b_1$ and $P_{H2} = P_{V2} = 1/2$. As a result, the above expressions reduce to
\begin{eqnarray}
    \avgpn{\hat{R}_{H}} & = & \left| b_1 \right|^2 \left(1 +  |T_H| \sqrt{P_H} \cos (\phi)\right), \\
    \avgpn{\hat{R}_{V}} & = & \left| b_1 \right|^2 \left(1 + \I  |T_V| \sqrt{P_V} \cos (\phi + \theta - \xi)\right),
\end{eqnarray}
with visibilities
\begin{eqnarray}
    \vis{H} & = & |T_H|\sqrt{P_H}, \label{eq:visHs} \\
    \vis{V} & = & \I |T_V|\sqrt{P_V}. \label{eq:visVs} 
\end{eqnarray}

\section{Post-interaction idler state}

The post-interaction state of the idler photon is
\begin{align}
    \rho_I^{\mathrm{post}} = \frac{1}{2} \ket{\psi_{Ib'}}\bra{\psi_{Ib'}} + \frac{1}{2} \left( \ident / 2 \right).
    \label{eq:poststate}
\end{align}
The final state is a uniform mixture of the maximally mixed state $\ident/2$ and the original pure state $\ket{\psi_{I}}$ in path $b'$. Evidently, the post-interaction state of the idler photon still contains some information about its original state of polarization.

\section{Calibration}

\begin{figure}[htbp]
\centering
\includegraphics[width=1\linewidth]{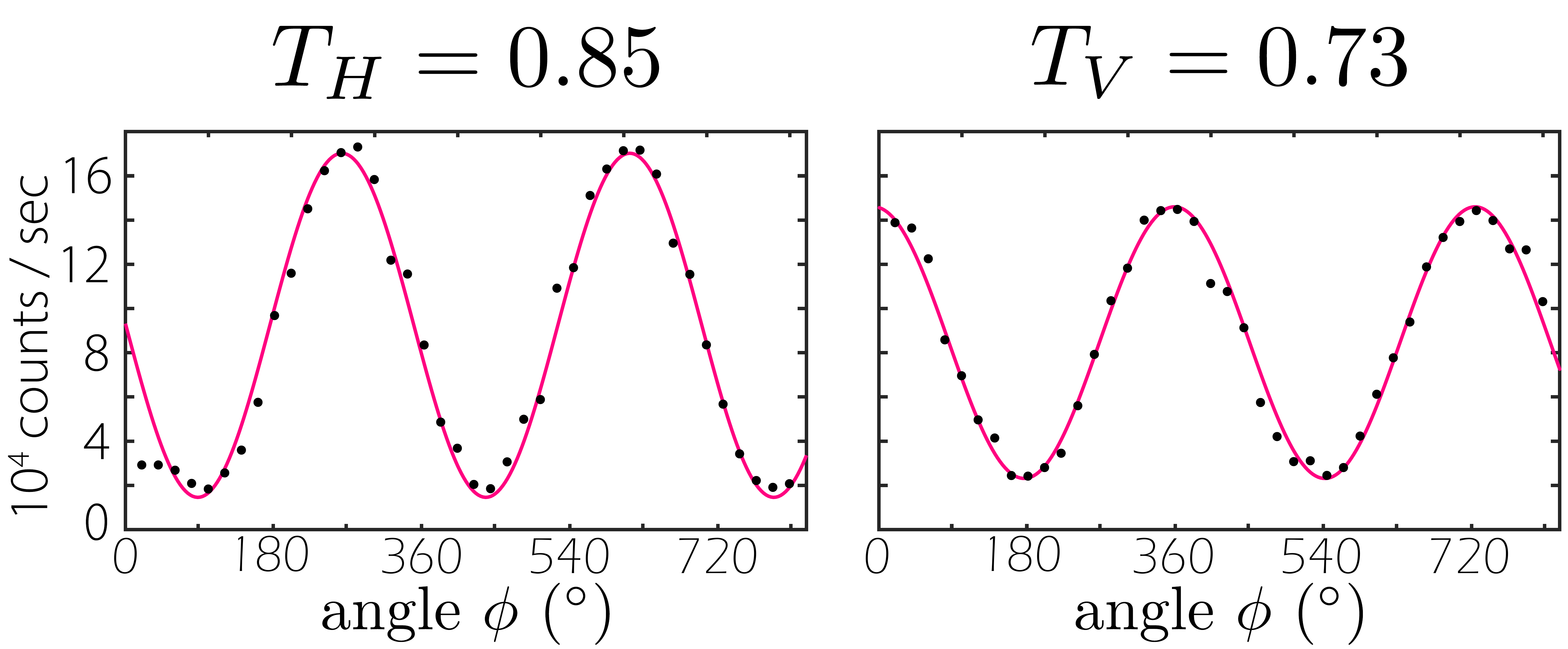}
\caption{Experimental visibility calibration. The maximum visibility $|T_H| = 0.85 \pm 0.03$ is obtained from the intensity profile measured by the detector D1 when the idler photon is in the state $\ket{H_{Ib'}}$ and the signal photon is in the state $\ket{H_{Sa}}$. The maximum visibility $|T_V| = 0.73 \pm 0.02$ is obtained for the setting $\ket{V_{Ib'}}$ and $\ket{V_{Sa}}$, when the intensity profile is recorded by the detector D2. Errors are given by the standard deviation of a sinusoidal function fitted to the experimental data.}
\label{Fig:visibilities-h-v}
\end{figure}

In the real experiment, the values of visibilities have to be calibrated first as the maximum achievable visibility may not be unity due to technical imperfections of the experimental setup, cf. Eqs.~(6) and (7) in the main text. The calibration measurement with maximum visibilities for $H$ and $V$ polarizations is presented in Fig.~\ref{Fig:visibilities-h-v}.

%\newpage

\section{Density matrix reconstruction by maximum-likelihood estimation technique}
To reconstruct the density matrix of the undetected photon we employ the maximum-likelihood-reconstruction technique \cite{PhysRevA.55.R1561} that consists in least-square fitting of a parametrized valid density matrix to the experimental data \cite{PhysRevA.64.052312}. At first, keeping the polarization of the idler photon fixed we collect the experimental data in the form of detector counts for each setting of the half-wave plate HWP$_S$ and phase shift $\phi$, see Fig.~1 in the main text. The resulting data $\{ h_k \}_k$ and $\{ v_k \}_k$ for horizontal and vertical polarization of the signal photon, respectively, should follow Eqs. (4) and (5) in the main text. We construct a cost function in the form
\begin{multline}
     f(\{ h_k \}_k, \{ v_k \}_k, P_H, \xi, \I) = \\ \sum_k (n \avgpn{\hat{R}_{H}} - h_k)^2 + \sum_k (n \avgpn{\hat{R}_{V}} - v_k)^2,
\end{multline}
where $n$ stands for the total number of counts for a fixed setting of HWP$_S$, expression $\avgpn{\hat{R}_{H}}$ is a function of $P_H$ and $\phi_k$, and $\avgpn{\hat{R}_{V}}$ is a function of $1 - P_H$, $\phi_k$, $\xi$, and $\I$. Function $f$ represents the difference between the experiment and the theoretical prediction. Its minimum is obtained for the optimal fitting parameters $P_H$, $P_V = 1 - P_H$, $\xi$, and $\I$. Using minimization routines such as \texttt{FindMinimum} in \textit{Mathematica} we find the optimal values of these parameters, which are then plugged into Eq.~(1) in the main text to reconstruct the corresponding density matrix.

\section{Fidelities}
\begin{figure}[htbp]
    \centering
    \includegraphics[width=1\linewidth]{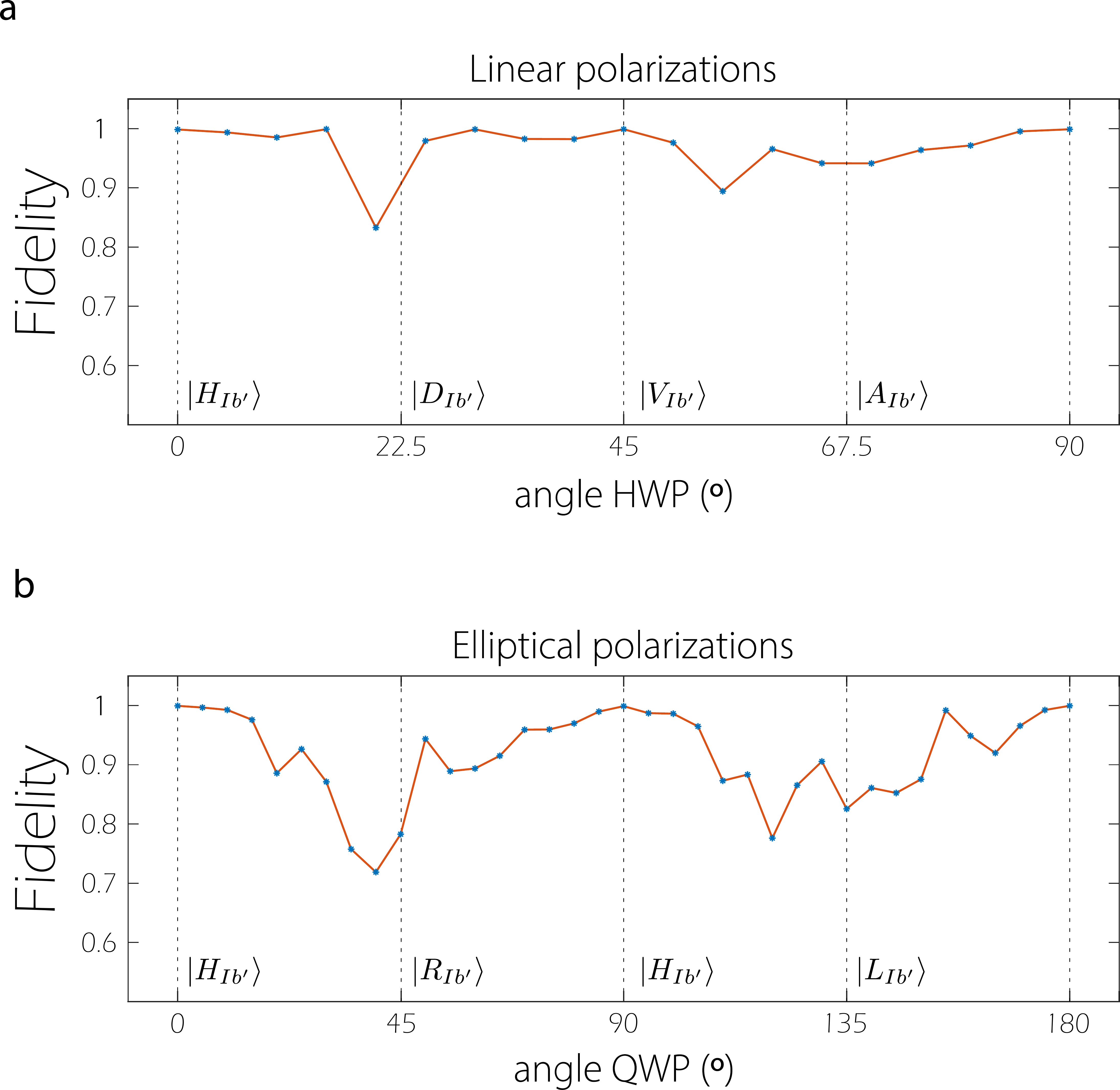}
    \caption{Fidelities of reconstructed states. (a) The fidelities for a HWP with the mean and lowest fidelities being $\Bar{F}_\mathrm{HWP}= 0.97$ and $F_\mathrm{HWP}^{\mathrm{(min)}}= 0.83$, respectively. (b) The fidelities for a QWP with the mean and lowest fidelities being $\Bar{F}_\mathrm{QWP}= 0.92$ and $F_\mathrm{QWP}^{\mathrm{(min)}}= 0.72$, respectively.}
    \label{fig:fids}
\end{figure}
The fidelity of each state is calculated by 
\begin{align}
    F=|\bra{\psi_{th}} \psi_{ex} \rangle|^2,
\end{align}
with $\ket{\psi_{ th}}$ the theoretical prediction, and $\ket{\psi_{ex}}$ the experimental result. In Fig.~\ref{fig:fids} the fidelities for the linearly polarized as well as elliptically polarized states are shown. As follows from the plot, the worst-case scenarios correspond to states close to $\ket{D_{I b'}}$ and $\ket{R_{I b'}}$, respectively.
%\bibliography{ref}
%\newpage